\documentclass[conference,a4paper]{IEEEtran}
\IEEEoverridecommandlockouts

\usepackage{cite}
\usepackage{amsmath,amssymb,amsfonts}
\usepackage{algorithmic}
\usepackage{graphicx}
\usepackage{textcomp}
\usepackage{xcolor}
\usepackage{siunitx}

\usepackage{tikz}
\usepackage{tikz-3dplot}
\usetikzlibrary{arrows.meta, 3d,  perspective, shapes, positioning, fit, calc, math,shapes.multipart}
\usepackage{pgfplotstable}
\usepgfplotslibrary{colormaps}
\pgfplotsset{compat=1.18}
\usepgfplotslibrary{units,statistics}

\usepackage[colorlinks=true, urlcolor=blue, citecolor=blue, linkcolor=darkgray]{hyperref}

\def\BibTeX{{\rm B\kern-.05em{\sc i\kern-.025em b}\kern-.08em
    T\kern-.1667em\lower.7ex\hbox{E}\kern-.125emX}}

\usepackage[nolist,nohyperlinks]{acronym}
\begin{acronym}
\acro{MMR}{magneto-mechanical resonator}
\end{acronym}

\usepackage{color}
\definecolor{ibilight}{RGB}{193,216,237}
\definecolor{ibidark}{RGB}{0,73,146}	
\definecolor{uke2}{RGB}{170,156,143} 	
\definecolor{uke3}{RGB}{87,87,86}		
\definecolor{ukesec1}{RGB}{255,223,0}	
\definecolor{ukesec2}{RGB}{239,123,5}	
\definecolor{ukesec3}{RGB}{104,195,205}	
\definecolor{ukesec4}{RGB}{138,189,36}	
\definecolor{ukesec5}{RGB}{178,34,41}	
\definecolor{tuhh}{RGB}{45,198,214}    

\definecolor{ibidarkBG}{RGB}{227,229,242}  
\definecolor{uke2BG}{RGB}{233,228,225} 	  
\definecolor{uke3BG}{RGB}{230,231,232}	    
\definecolor{ukesec1BG}{RGB}{255,243,190}   
\definecolor{ukesec2BG}{RGB}{254,232,212}   
\definecolor{ukesec3BG}{RGB}{222,241,241}   
\definecolor{ukesec4BG}{RGB}{233,243,222}  
\definecolor{ukesec5BG}{RGB}{244,230,225}

\begin{document}
\title{Phase-Matched Chirps for Energy-Efficient Excitation of Magneto-Mechanical Resonators
\thanks{This project is funded by the Deutsche Forschungsgemeinschaft (DFG, German Research Foundation) - SFB 1615 - 503850735 and 555456057.}
}

\author{

\IEEEauthorblockN{Sarah Reiss\IEEEauthorrefmark{1}\IEEEauthorrefmark{2}, Fabian Mohn\IEEEauthorrefmark{1}\IEEEauthorrefmark{2}, Jonas Faltinath\IEEEauthorrefmark{1}\IEEEauthorrefmark{2}, and Tobias Knopp\IEEEauthorrefmark{1}\IEEEauthorrefmark{2}\IEEEauthorrefmark{3}\\
E-mail: sarah.reiss@tuhh.de}

\IEEEauthorblockA{\IEEEauthorrefmark{1}Institute for Biomedical Imaging, Hamburg University of Technology, Hamburg, Germany\\
\IEEEauthorrefmark{2}Section for Biomedical Imaging, University Medical Center Hamburg-Eppendorf, Hamburg, Germany\\
\IEEEauthorrefmark{3}Fraunhofer Research Institution for Individualized Medical Technology and Engineering IMTE, Lübeck, Germany}
}

\newcommand{\IR}{\mathbb{R}}	
\newcommand{\phimax}{\varphi_\mathrm{max}}		

\maketitle

\begin{abstract}
Magneto-mechanical resonators (MMRs) are a recently proposed class of passive, wireless devices used for sensing and tracking. The operating principle of MMRs relies on an excitation of the sensor by oscillating magnetic fields, followed by an inductive readout to measure the sensor's dynamics. One challenge lies in generating efficient excitation signals to ensure maximum energy harvesting by the sensor. In this simulation study, we focus on chirped magnetic-field excitations and compare their performance with conventional constant-frequency excitation for both sinusoidal and rectangular waveforms. The results indicate varying levels of efficiency, with a sinusoidal linear down-chirp excitation signal yielding the largest deflection angle, offering the best potential for reducing the energy consumption and improving the temporal resolution of the measurement.
\end{abstract}

\begin{IEEEkeywords}
magneto-mechanical resonator (MMR), resonance frequency, passive sensing, wireless sensing, magnetic localization, magnetic sensing, chirp
\end{IEEEkeywords}

\section{Introduction}
Mechanical resonators periodically exchange kinetic and potential energy, sustaining an oscillation. When driven by an external force near resonance, energy is transferred into the system at a rate governed by its quality factor $Q$, leading to an increasing oscillation amplitude \cite{shabana2019theoryVibration}.

\Acp{MMR} are a certain type of resonators that exhibit damped oscillations following external magnetic excitation. They consist of two permanent magnets in an antiparallel arrangement: a stator fixed to the housing, and a rotor suspended by a thin filament~(see~\autoref{fig:MMR}). Gleich et al.~\cite{Gleich2023Miniature} recently introduced them as a passive, wireless sensing and tracking tool, with future applications in medical engineering e.g. for guidance in navigating surgical devices during minimal invasive procedures~\cite{Gleich2023Miniature} or in process engineering, where MMRs enable simultaneous measurement of key process parameters and their spatial distribution~\cite{Merbach2025Wireless}.

A time-varying external magnetic field exerts a periodic torque on the rotor, deflecting it from its equilibrium position and initiating torsional oscillation about its filament axis.
The instantaneous frequency of the MMR decreases non-linearly with rising instantaneous oscillation amplitude~\cite{Gleich2023Miniature, faltinath_natural_2025}. 
With static excitation frequencies used in previous MMR studies~\cite{Gleich2023Miniature, faltinath_natural_2025,reiss2026ParamEstim,Merbach2025Wireless, mohn2025TxTxConcepts,knopp2024emperical}, the inevitable phase drift between the transmit signal and the decelerating sensor causes destructive interference, actively suppressing rotor motion and ultimately canceling the signal.
This motivates a design of the excitation waveform that is precisely matched to the instantaneous frequency and phase of the \ac{MMR} and maximizes torque. 

The purpose of this study is to compare different waveforms, sinusoidal and rectangular, and instantaneous phase models -- fixed frequency and linear chirp -- to investigate their efficiency based on the maximum deflection angle achieved after excitation. To ensure a fair comparison, all excitation signals are normalized to the same input energy.

\section{Materials and Methods}

\begin{figure}
\centering
    \newcommand{\zylinderTop}{5.2} 
\newcommand{\zylinderBot}{0}  
\newcommand{\distTopRotor}{2} 
\newcommand{\distBotStator}{1.1}  
\newcommand{\heightFix}{0.5}
\newcommand{\colorMagOne}{ibidark} 
\newcommand{\colorMagTwo}{ukesec5} 
\newcommand{\shadeRadius}{0.186} 
\newcommand{\magRadius}{0.9} 

\begin{tikzpicture}[scale = 0.8]

  \def\R{1.2}           
  \def\dy{2.5*\R}       
  \def\mMag{1.6*\R}      
  \def\gap{0.4*\R}      
  \def\housingR{1.2*\R}  
  \def\phi{40}           
  \def\T{10}             
  
\node[] (initNode) at (0,0) {};
\begin{scope}[shift={(initNode.center)},scale=1]
  \begin{axis}[
    view={-60}{10},
    axis equal image,
    hide axis,
    samples=40,
    domain=0:80,
    y domain=0:360,
    anchor = center,
    clip = false
  ]

  \addplot3 [
    surf,
    opacity=0.1,
    z buffer=sort,
    domain=0:2*pi,
    y domain=\zylinderBot : \zylinderTop,
    colormap={split}{color=(black!8!white); color=(black!8!white)},
    shader=interp,
    draw=none,
  ]
  ({cos(deg(x))}, {sin(deg(x))}, y);

  \addplot3 [black, thick,  samples=5, domain=\zylinderTop - \distTopRotor+\magRadius : \zylinderTop] (0,0,{x});

     \addplot3 [
    surf,
    z buffer=sort,
    domain=-50:120,
    y domain=\zylinderBot : \zylinderBot+\heightFix,
    colormap={split}{color=(black!30!white); color=(black!8!white); color=(black!45!white)},
    shader=interp,
    draw=none,
  ]
  ({cos(x)}, {sin(x)}, y);

  \addplot3 [
    surf,
    shader=interp,
    z buffer=sort,
    domain=0:180,
    y domain=0:180,
    colormap={split}{color=(\colorMagOne); color=(\colorMagOne)}
  ]
  ({\magRadius*sin(x)*cos(y)}, {\magRadius*sin(x)*sin(y)}, {\magRadius*cos(x) + \zylinderTop -\distTopRotor});
  \addplot3 [
    surf,
    shader=interp,
    z buffer=sort,
    domain=0:180,
    y domain=180:360,
    colormap={split}{color=(\colorMagTwo); color=(\colorMagTwo)}
  ]
  ({\magRadius*sin(x)*cos(y)}, {\magRadius*sin(x)*sin(y)}, {\magRadius*cos(x) + \zylinderTop -\distTopRotor});
  \shade[ball color=white, opacity=0.3] (axis cs: 0,0,\zylinderTop -\distTopRotor) circle [radius= \shadeRadius, x = 1, y = 1, z= 1];
  \shade[ball color=gray!20, opacity=0.2] (axis cs: 0,0,\zylinderTop -\distTopRotor) circle [radius= \shadeRadius, x = 1, y = 1, z= 1];
    
  \addplot3 [
    surf,
    shader=interp,
    z buffer=sort,
    domain=0:180,
    y domain=0:180,
    colormap={split}{color=(\colorMagTwo); color=(\colorMagTwo)}
  ]
  ({\magRadius*sin(x)*cos(y)}, {\magRadius*sin(x)*sin(y)}, {\magRadius*cos(x)+ \distBotStator +\zylinderBot});
  \addplot3 [
    surf,
    shader=interp,
    z buffer=sort,
    domain=0:180,
    y domain=180:360,
    colormap={split}{color=(\colorMagOne); color=(\colorMagOne)}
  ]
  ({\magRadius*sin(x)*cos(y)}, {\magRadius*sin(x)*sin(y)}, {\magRadius*cos(x)+ \distBotStator+\zylinderBot});
  \shade[ball color=white, opacity=0.3] (axis cs: 0,0,\distBotStator+\zylinderBot) circle [radius= \shadeRadius, x = 1, y = 1, z= 1];
  \shade[ball color=gray!20, opacity=0.2] (axis cs: 0,0,\distBotStator+\zylinderBot) circle [radius= \shadeRadius, x = 1, y = 1, z= 1];

   \addplot3 [
    surf,
    z buffer=sort,
    domain=120:310,
    y domain=\zylinderBot : \zylinderBot+\heightFix,
    colormap={split}{color=(black!30!white); color=(black!8!white); color=(black!45!white)},
    shader=interp,
    draw=none,
  ]
  ({cos(x)}, {sin(x)}, y);

  \addplot3 [
    surf,
    z buffer=sort,
    domain=0:2*pi,
    y domain=\zylinderTop-\heightFix : \zylinderTop,
    colormap={split}{color=(black!30!white); color=(black!8!white); color=(black!45!white)},
    shader=interp,
    shading=axis,
    draw=none,
    fill = gray!20,
  ]
  ({cos(deg(x))}, {sin(deg(x))}, y);
  \begin{scope}[shift={(axis cs:-1.5,1,0)}]
    \coordinate (O) at (axis cs:0,0,0);

    \draw[thick,-{Stealth[inset=0pt, length=2mm]}] (O) -- (axis cs:1,0,0.0) node[anchor=south east, inner sep = 2 ] {$v$};
    \draw[thick,-{Stealth[inset=0pt, length=2mm]}] (O) -- (axis cs:0,-1,0) node[anchor=north, inner sep = 2] {$u$};
    \draw[thick,-{Stealth[inset=0pt, length=2mm]}] (O) -- (axis cs:0,0,1) node[anchor=south east, inner sep = 0] {$w$};
  \end{scope}
  \end{axis}
\end{scope}

\node[anchor = center] at  ($(initNode.center)+(3,1.7)$) {\textbf{top view}};
\begin{scope}[shift={($(initNode.center)+(3,-0.3)$)}]
    
    \draw[thick] (0,0) circle (\R);
    \begin{scope}[rotate=\phi]
      \clip (0,0) circle (\R);
      \fill[ibidark] (-\R,-\R) rectangle (0,\R);  
      \fill[ukesec5] (0,-\R) rectangle (\R,\R);   
    \end{scope}
    \shade[ball color=white, opacity=0.3] (0,0) circle (\R);
    \shade[ball color=gray!20, opacity=0.2] (0,0) circle (\R);
    
    \draw[-{Stealth[inset=0pt, length=2mm]}, thick] (0,0) -- ({1*\mMag*cos(\phi)},{1*\mMag*sin(\phi)})
    node[anchor = north west] {$\pmb{m}$};

    \draw[dashed,-{Stealth[inset=0pt, length=2mm]}, thick] (0,0) -- (\mMag,0)  node[anchor = north] {$\pmb{m}_\mathrm{eq}$};

    \draw[{Stealth[inset=0pt, length=2mm]}-{Stealth[inset=0pt, length=2mm]}, thick] ({1.6*\mMag*cos(\phi)},{-1*\mMag*sin(\phi)}) -- ({1.6*\mMag*cos(\phi)},{1*\mMag*sin(\phi)});
    \node[anchor = west] at ({1.6*\mMag*cos(\phi)},0) {$\pmb{B}_\mathrm{ext}$};
    
    \begin{scope}[thick,-{Stealth[inset=0pt, length=2mm]}]
      \draw (0,0) -- (0.5,0) node[anchor=north west, inner sep = 2] {$\pmb{u}$};
      \draw (0,0) -- (0,0.5) node[anchor=south east, inner sep = 2] {$\pmb{v}$};
      \draw[thick, fill] (0,0) circle (0.05); 
      \fill (0,0) circle (0.5pt);
      \node[anchor=east] at (0,0) {$\pmb{w}$};
    \end{scope}
    
    \draw[thick,-] (0.7,0) arc[start angle=0,end angle=\phi,radius=0.7];
    \node[anchor=south] at (0.5,-0.05) {$\varphi$};
\end{scope}

\end{tikzpicture}
    \caption{\textbf{Left:} The MMR consists of two permanent magnets in equilibrium position within the local $uvw$-coordinate system. 
    \textbf{Right:} The rotor's magnetic moment $\pmb{m}$ is deflected, quantified by the deflection angle $\varphi$, from the equilibrium position $\pmb{m}_\mathrm{eq}$ due to an external magnetic field $\pmb{B}_\mathrm{ext}$ in $v$-direction.}
    \label{fig:MMR}
\end{figure}
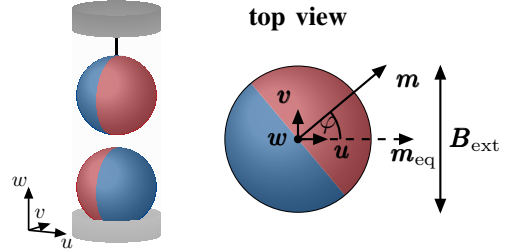

An \ac{MMR} measurement system consists of an \ac{MMR}, an inductive coil array for remote excitation and readout, and a signal-processing platform for real-time analysis and excitation control. Measurements are divided in several frames comprising a TX and an RX window. During TX, the coil array applies an excitation signal $B_\mathrm{ext}$, whereas during RX $B_\mathrm{ext}$ is switched off and the damped oscillation of the rotor's magnetic moment is detected inductively. The rotor’s equation of motion \cite{Gleich2023Miniature} is
\begin{equation}
    \ddot{\varphi}(t)+\frac{\omega_{\text{nat}}}{Q} \dot{\varphi}(t) +\omega_{\text{nat}}^2 \sin(\varphi(t))= \frac{T_d(t)}{I}
    \label{eq:ODE}
\end{equation}
with the deflection angle $\varphi(t)$ relative to equilibrium position, the natural frequency $\omega_{\text{nat}}$, and the rotor's moment of inertia $I$. 
The excitation field, orthogonal to the rotor's magnetic moment $m$ in equilibrium position (see \autoref{fig:MMR}), generates a driving torque $T_d(t)$, depending on $\varphi$: $T_d(t) = m B_\mathrm{ext}(t) \cos (\varphi(t))$.

\subsection{Excitation Waveforms}
Two fundamental concepts for generating excitation signals in MMRs are based on either sinusoidal or rectangular waveforms. Sinusoidal excitation signals can be generated using commercially available Class-D audio amplifiers, while rectangular excitation pulses with variable duty cycles are typically produced by H-bridge circuits~\cite{mohn2025TxTxConcepts}.

A rectangular magnetic field signal $B_\mathrm{rect}(t)$ with amplitude $B_v$ and duty cycle $D = \tau/T_\mathrm{rect} \in [0,1]$ alternates between $+B_v$, $-B_v$, and zero, governed by the instantaneous phase $\psi(t)$, such that the fraction $D$ of each half-period carries a non-zero field value.
The sinusoidal magnetic field signal can be described by
\begin{equation}
    B_\mathrm{sin}(t) =B_v \,\sin\!\left(\psi(t)\right).
    \label{eq:B_fSin}
\end{equation}

Regarding the instantaneous phase $\psi(t)$, we analyze three different models. First, using the fixed frequency $\omega_\mathrm{RX,end}$ at the end of the pervious RX window, to be directly in resonance with the rotor's movement at the beginning of TX
\begin{equation}
    \psi_0(t) = \omega_\mathrm{RX,end}t + \psi_\mathrm{RX,end}.
\end{equation}
The initial phase $\psi_\mathrm{RX,end} \in [0,2\pi)$ is calculated using the last period of the previous RX window (identical for all three models). We call those signals F0Sin and F0Rec. 

Second, using the fixed frequency $\omega_\mathrm{avg}$ determined from the average of $\omega_\mathrm{RX,start}$ and $\omega_\mathrm{RX,end}$ (frequency at the beginning and the end of the RX window), in order to compensate for frequency changes occurring during the TX phase while preserving a straightforward realization of the excitation signal
\begin{equation}
    \psi_\mathrm{avg}(t) = \omega_\mathrm{avg}t + \psi_\mathrm{RX,end}.
\end{equation}
We call those signals FAvgSin and FAvgRec. 

Third, we propose a linear down-chirp as a more complex signal to match the frequency change with increasing deflection angle. The instantaneous phase is described as
\begin{equation}
    \psi_\mathrm{chirp}(t) =  \frac{c}{2}t^2 + \omega_\mathrm{RX,end}t
    + \psi_\mathrm{RX,end}
\end{equation}
with a linear frequency decrease during TX. The slope $c = (\omega_\mathrm{RX,start}-\omega_\mathrm{RX,end})/T_\mathrm{TX}$ is acquired using the duration of the TX window $T_\mathrm{TX}$. We call those signals ChirpSin and ChirpRect. Exemplary signals for F0Sin, FAvgRect ($D = 0.7$) and ChirpSin with corresponding angular velocity are displayed in \autoref{fig:results_dutyCycle}, as these signals have to be in phase for efficient excitation.

\subsection{Energy}
To calculate the energy consumption over the complete measurement, we use the real power $P = RI_\mathrm{RMS}^2$ with the coil resistance $R$. Based on $I = B_v/p_v$, with $p_v$ denoting the coil sensitivity at the MMR position, the root mean square (RMS) current is given by $I_{\mathrm{RMS}} = I/\sqrt{2}$ for sinusoidal and $I_{\mathrm{RMS}} = I\sqrt{D}$ for rectangular signals. For the signals using fixed excitation frequencies the energy can be calculate to be
\begin{equation}
    E  = T_\mathrm{TX} N \cdot RI_\mathrm{RMS}^2 
    \label{eq:energy_others}
\end{equation}
where $N$ is the number of frames. 
Because the frequency of ChirpSin and  ChirpRect changes dynamically, the energy is calculated using
\begin{equation}
    E_\mathrm{chirp} = R \sum_{i=1}^{N} \int_0^{T_\mathrm{TX}} I_\mathrm{chirp, i}(t)^2 \,dt ,
    \label{eq:energy_chirp}
\end{equation}
where the current $I_\mathrm{chirp, i}$ depends on each frame due to an individual slop.

\subsection{Simulation}
The combination of two waveforms and three instantaneous phases results in six different excitation signals, which are investigated via Equation~\eqref{eq:ODE} in a simulation framework using Julia and is published under the MIT License at \url{https://github.com/IBIResearch/MMR-Phase-Matched-Excitation}.

The coil arrangement in Helmholtz configuration and the smaller MMR used in \cite{reiss2026ParamEstim} with $\omega_\mathrm{nat} = 2\pi \cdot \SI{814}{\per\second}$, $Q$ = 852, a rotor diameter of \SI{1}{mm}, and a remanence of \SI{1.44}{T} are implemented. The MMR is positioned in the center ($\pmb{r} =\pmb{0}$) and oriented such that the $v$-direction was aligned with the coil axis of the $x$ coil. Hence, only the coil sensitivity of the $x$ coil in $x$ direction at $p_x (\pmb{0}) = p_v(\pmb{0}) = \SI{879.5}{\micro \tesla/A}$ and the coil resistance $R_x= \SI{4.0}{\ohm}$ are needed. The amplitude of the magnetic field for ChirpSin is set to \SI{40}{\micro \tesla}, while the amplitudes for the other methods are adjusted to maintain the same energy consumption.

First different duty cycles $D \in \{0.4, 0.5, \ldots, 1.0\}$ were tested with FAvgRect, and the one achieving the highest $\phimax$ chosen for the following experiments. Subsequently, 100 simulations were conducted for different TX durations, uniformly sampling the parameter uncertainties of $\pm 2\pi \cdot \SI{1}{\per\second}$ for $\omega_\mathrm{RX,start}, \omega_\mathrm{RX,end}$ and $\pm \SI{10}{\degree}$ for  $\psi_\mathrm{RX,end}$.

\section{Results}
\begin{figure}
    \centering
    \includegraphics[width=\columnwidth]{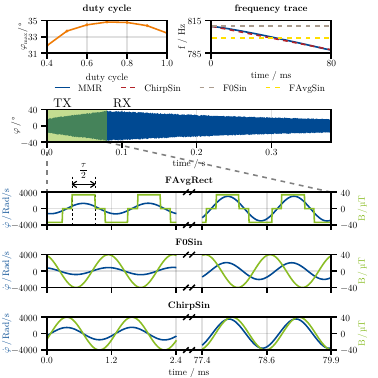}
    \caption{\textbf{Upper-left:} Maximum deflection angle for different duty cycles for rectangular excitation (FAvgRect). \textbf{Upper-right:} Frequency traces of MMRs deflection angle and excitation signals. \textbf{Bottom:} Deflection angle of an exemplary MMR frame (TX/RX window). Zooms show a few periods of excitations signals with corresponding angular velocity. }
    \label{fig:results_dutyCycle}
\end{figure}

\autoref{fig:results_dutyCycle} on the upper left compares the maximum deflection angle $\phimax$ across different duty cycles, revealing a monotonically increasing response up to an optimal duty cycle of $D = 0.7$, beyond which $\phimax$ decreases gradually.

The expected decrease of the MMR's instantaneous frequency during TX results in increasing frequency differences for F0Sin as time progresses within the TX window~(see \autoref{fig:results_dutyCycle}). This observation is consistent with the phase mismatch towards the end of the TX window, yielding a mean phase difference of $\Delta\psi = 44.7^\circ \pm 39.5^\circ$ over the entire TX interval. For FAvgRect, the excitation signal is approximately in phase with the angular velocity at the beginning and end of the TX window; however, the highest phase lead is observed in the middle, resulting in $\Delta\psi = -29.22^\circ \pm 13.75^\circ$. ChirpSin exhibits the smallest phase difference, with $\Delta\psi = -10.9^\circ \pm 6.3^\circ$. 

\begin{figure}
    \centering
    \includegraphics[width=\columnwidth]{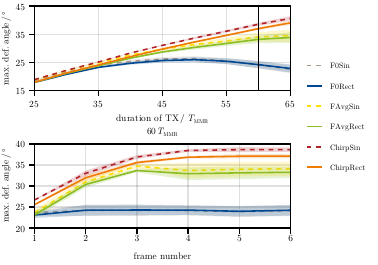}
    \caption{\textbf{Top:} Maximum deflection angle (mean and standard deviation) for different durations of TX for all methods and $B_\mathrm{ChirpSin} = \SI{40}{uT}$. \textbf{Bottom:} Progression of maximum deflection angle (mean and standard deviation) during one simulation experiment with $T_\mathrm{TX} = 60\,T_\mathrm{MMR}$ (vertical line in top plot).}
    \label{fig:results_TX}
\end{figure}

\autoref{fig:results_TX} compares the maximum deflection angle achieved by all methods for different TX window durations. The durations are defined as multiples of the MMR oscillation period ($T_\mathrm{MMR}= 2 \pi/\omega_\mathrm{nat}$). The respective sinusoidal excitation signals attain higher $\phimax$ than their corresponding rectangular counter part. All methods reach similar $\phimax$ for short TX windows. However, with increasing $T_\mathrm{TX}$, $\phimax$ increases the most for the chirp excitation signals. F0Sin and F0Rect reach a peak around $50\,T_\mathrm{MMR}$ and decrease afterwards.

During the simulation experiment with a TX window duration of TX $= 60\,T_\mathrm{MMR}$ all excitation methods exhibit a transient build-up phase before reaching the maximum achievable deflection angle. Although the down-chirp excitation takes the longest to reach a steady-state $\phimax$, it achieves higher deflection angles more rapidly and the highest maximum $\phimax$ compared to all other excitation methods.

\section{Discussion and Conclusion}
By improving the phase alignment between the excitation signal and the resonator dynamics, we increased the energy transferred to the system using a linear down-chirp sinusoidal excitation signal, resulting in faster excitation and higher values of $\phimax$. For $T_\mathrm{TX} = 60\,T_\mathrm{MMR}$ the maximum deflection angle $\phimax$ is increased by \SI{14}{\percent} compared to the best excitation signal with fixed frequency. 

The instantaneous phase depends on the corresponding instantaneous frequency. In case that the instantaneous frequency of the excitation signal remains fixed while the MMR frequency decreases, the phase mismatch increases over time. As the phase difference approaches \SI{90}{\degree}, the driving torque acts against the rotor's motion, leading to signal destruction and reduced performance. We observe this effect starting with a TX window duration of $45$ MMR oscillation periods for F0Sin and F0Rect. While using the average frequency provides a reasonable approximation, a linear down-chirp remains largely in resonance with the rotor motion throughout the TX window, thereby reinforcing the oscillation over the entire excitation period. Although the actual frequency decreases nonlinearly, the down-chirp remains largely in phase with the rotor motion and is therefore well-suited for phase-matched excitation. 

For rectangular waveforms, we observe that a duty cycle of $0.7$ is most efficient. $\phimax$ is smaller for short, energy-normalized excitation pulses due to a broader frequency spectrum for shorter pulses, redistributing energy from the resonance to higher harmonics.
However, excessively long duty cycles are also less efficient, as the torque depends on the angle of deflection via a cosine relationship, it decreases as the deflection angle increases. In addition, phase mismatches are particularly critical for rectangular excitation signals, as the excitation may already apply a torque in the opposite direction to the rotor's motion near the turning points.

In summary, the linear down-chirp is a promising excitation method for MMRs. The increased energy transfer enables faster excitation and higher $\phimax$, allowing the target amplitude to be reached within shorter TX windows and thereby improving the effective temporal resolution, since no measurements can be acquired during TX. The linear down-chirp was chosen due to its simplicity, introducing only a single additional parameter, the chirp slope, which can be estimated using only the previous RX window and adapted to different TX durations. 

While effective, further optimization for specific sensors and applications remains possible, for example by using a steeper initial slope during the start-up phase to accelerate convergence or to achieve the desired amplitude within a single excitation window.
Experimental validation is an important next step, for example by adapting the signal generator in the setup presented in \cite{reiss2026ParamEstim} to enable the generation of chirp signals.

\pagebreak

\bibliography{ref} 

\end{document}